# Adaptive convolutional neural networks for *k*-space data interpolation in fast magnetic resonance imaging

Tianming Du[1,2], Honggang Zhang[1], Yuemeng Li[2], Hee Kwon Song[2], Yong Fan[2]

[1] School of Information and Communication Engineering, Beijing University of Posts and Telecommunications, Beijing, 100876, China
[2] Department of Radiology, Perelman School of Medicine, University of Pennsylvania, Philadelphia, PA 19104, USA

**Abstract.** Deep learning in *k-space* has demonstrated great potential for image reconstruction from undersampled *k*-space data in fast magnetic resonance imaging (MRI). However, existing deep learning-based image reconstruction methods typically apply weight-sharing convolutional neural networks (CNNs) to *k*-space data without taking into consideration the *k*-space data's spatial frequency properties, leading to ineffective learning of the image reconstruction models. Moreover, complementary information of spatially adjacent slices is often ignored in existing deep learning methods. To overcome such limitations, we have developed a deep learning algorithm, referred to as adaptive convolutional neural networks for *k*-space data interpolation (ACNN-k-Space), which adopts a residual Encoder-Decoder network architecture to interpolate the undersampled k-space data by integrating spatially contiguous slices as multi-channel input, along with *k*-space data from multiple coils if available. The network is enhanced by self-attention layers to adaptively focus on *k*-space data at different spatial frequencies and channels. We have evaluated our method on two public datasets and compared it with state-of-the-art existing methods. Ablation studies and experimental results demonstrate that our method effectively reconstructs images from undersampled *k*-space data and achieves significantly better image reconstruction performance than current state-of-the-art techniques.

## 1 Introduction

Deep learning has shown great potential for image reconstruction from undersampled *k*-space data in fast magnetic resonance imaging (MRI) [1, 2]. A variety of deep learning methods have been developed to solve the image reconstruction problem, including model-driven [3-10] and data-driven methods [11-19]. Unlike model-driven deep learning methods whose performance is hinged on their model capacity, data-driven methods directly learn a mapping between undersampled *k*-space data and reconstructed images [15], or an interpolation in either image domain [11-14] or *k*-space [16-19]. Particularly, fully connected neural networks have been used to learn the Fourier transform itself [15]. However, it is difficult to use such a method to reconstruct large size images due to huge memory requirements of fully connected neural networks. In contrast, convolutional neural networks (CNNs) with weight sharing are memory efficient and therefore have been widely adopted to learn an interpolation from undersampled *k*-space data for image reconstruction in conjunction with Fast Fourier transform (FFT) [16-19].

Since CNNs in *k*-space could be used to directly interpolate the missing *k*-space samples [16-19], it is reasonable to believe that they could perform better than their counterparts working in the image domain with the same network architecture. However, existing *k*-space deep learning methods directly adopt CNNs without taking into consideration characteristics of the *k*-space data. First, the samples at the central *k*-space region (low spatial frequencies) contain most of the information of image contrast, while the samples further away from the center (high spatial frequencies) contain information about the edges and boundaries of the image. Therefore, applying weight-sharing CNNs to the entire *k*-space data, as used in existing *k*-space deep learning methods, ignores distinctive contributions of different spatial frequencies of the *k*-space data to the image reconstruction, leading to ineffective learning of CNNs. Second, undersampled *k*-space data of spatially adjacent image slices may provide complementary information for image reconstruction. However, most existing deep learning methods typically learn interpolations for spatially adjacent image slices independently, ignoring their complementary information that could potentially improve image reconstruction, except two recent methods [9, 20]. Third, in multi-coil MRI acquisition, *k*-space data from different coils are sensitive to different regions of the object but are often treated equally as multiple channels in existing *k*-space deep learning image reconstruction methods, which may degrade image reconstruction performance.

In order to overcome the aforementioned limitations, we have developed a novel *k*-space deep learning framework for image reconstruction from undersampled *k*-space data, referred to as adaptive convolutional neural networks for

*k*-space data interpolation (ACNN-k-Space). Particularly, a residual Encoder-Decoder network architecture is adopted to interpolate the undersampled *k*-space data with CNNs enhanced by a self-attention layer [21], referred to as frequency-attention layer, which adaptively assigns weights to features learned by CNNs for *k*-space samples at different spatial frequencies. Moreover, instead of learning interpolations for spatially adjacent image slices independently, our method learns an interpolation for each image slice by integrating slices within its spatial neighborhood as a multi-channel input, along with *k*-space data from multiple coils if available, to the residual network. Since the image slices may contribute to the image reconstruction differently and data from different coils are sensitive to different regions of the object, we adopt another self-attention layer, referred to as channel-attention layer [22], to adaptively assigns weights to features learned by CNNs for different channels. Together, the residual Encoder-Decoder network with frequency-attention and channel-attention layers learns an interpolation for undersampled *k*-space data and reconstructs an image in conjunction with inverse FFT (IFFT) in an end-to-end fashion. We have evaluated our method based on two publicly available datasets. Ablation studies and experimental results show that our method could effectively reconstruct images from undersampled *k*-space data and achieve better image reconstruction performance than existing state-of-the-art techniques.

## 2  Methods

To generate missing *k*-space samples, we adopt a residual Encoder-Decoder network to reconstruct images from undersampled *k*-space data, as illustrated in Fig. 1. The residual network learns an interpolation to reconstruct images in conjunction with IFFT from a multi-channel input that consists of undersampled *k*-space data from spatially adjacent slices and data from multiple coils if available. Its Encoder-Decoder component consists of CNNs, enhanced by frequency-attention and channel-attention layers. The network is optimized by minimizing the mean square error between the reconstructed image and its corresponding image obtained from fully sampled *k*-space data.

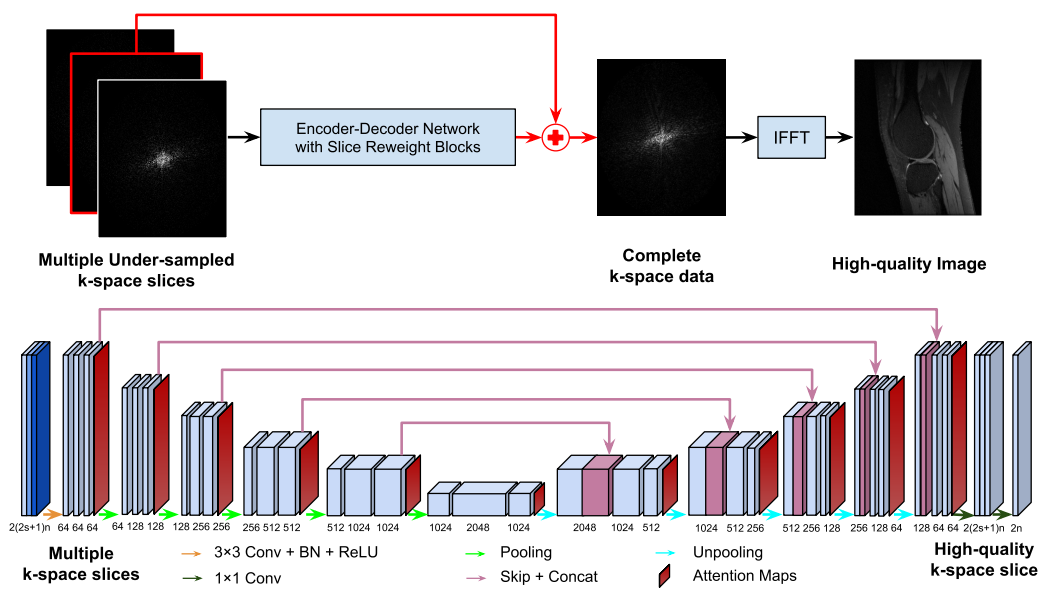

**Fig. 1.** A residual Encoder-Decoder network of CNNs, enhanced by frequency-attention and channel-attention layers, for image reconstruction from undersampled *k*-space data. The residual network (top row) learns an interpolation to reconstruct images in conjunction with IFFT from a multi-channel input that consists of undersampled *k*-space data from spatially adjacent slices and data from multiple coils if available. Its Encoder-Decoder component (bottom row) consists of multiple layers of CNNs, enhanced by frequency-attention and channel-attention layers. Attention maps are the outputs of the frequency-attention and channel-attention layers to modulate features learned by CNNs before every pooling or unpooling layer.

### 2.1 Encoder-Decoder network architecture

We adopt an Encoder-Decoder network to learn an interpolation to generate missing *k*-space samples for image reconstruction from undersampled *k*-space data. As illustrated in Fig. 1, the network backbone is a U-Net [23], consisting of convolutional layers, followed by rectified linear unit (ReLU) [24] and batch normalization (BN) [25], with parameters specified in Fig. 1 (bottom).

For image reconstruction of each slice from its undersampled $k$-space data, the network's input consists of complex valued $k$-space data of $c = 2s + 1$ slices within its spatial neighborhood, where $s \geq 0$ is the consecutive slices stacked on top and bottom of the slice under consideration. Complex values of the $k$-space data are split into two channels of real value signals. For imaging with multiple coils, $k$-space data of $n$ coils are stacked as multi-channel data with each coil's data as two channels of real value signals. Therefore, the number of channels of the input is $2(2s + 1)n$ and the number of channels of the output is $2n$ to form complex valued $k$-space data. For a slice with less than $s$ bottom or top slices, the data volume is padded with the first bottom slice or the first top slice, respectively.

In order to take into consideration distinctive contributions of different spatial frequencies and different channels to the image reconstruction, we adopt self-attention layers to enhance the learning of CNNs.

## 2.2 Attention layers

We adopt frequency-attention and channel-attention layers to enhance $k$-space deep learning. Both frequency-attention and channel-attention layers learn self-attention maps from their feature maps [21, 22] to modulate features learned by CNNs and the modulated feature maps are aggregated by an elementwise Max-out operation, as illustrated in Fig. 2.

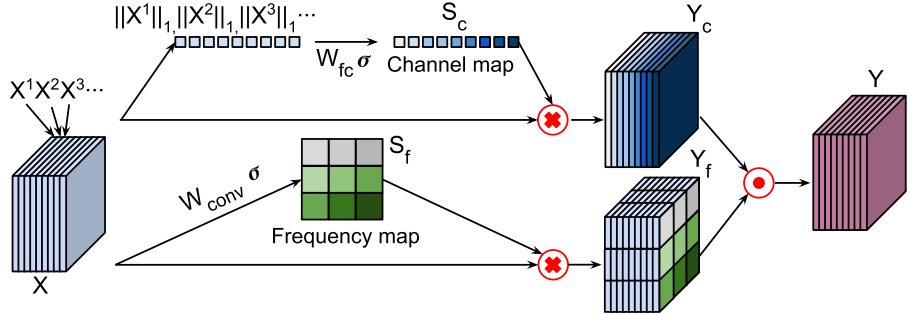

**Fig. 2.** Self-attention layers learn a frequency-attention map (bottom) and a channel-attention map (top) respectively, and weighted feature maps are aggregated by an elementwise Max-out operation. $X^i$ refers to the $i$th channel of the $k$-space feature maps, $Y_c$ refers to features weighted by the channel-attention map, $Y_f$ refers to features weighted by the frequency-attention map, and $Y$ refers to the aggregated features of $Y_c$ and $Y_f$ through a Max-out operation.

### 2.2.1 Frequency-attention layers

The $k$-space samples are in the spatial frequency domain, and the samples at low frequencies characterize most of the signal intensity and contrast information of the image, while the samples at high frequencies characterize information about objects' edges and boundaries. Existing $k$-space deep learning methods directly apply the weight-sharing CNNs to entire $k$-space data, ignoring distinct contributions of different frequencies of the $k$-space data to image reconstruction. To tackle this issue, we adopt a frequency-attention layer to modulate features learned by CNNs at different spatial frequencies. As illustrated in Fig. 2 (bottom), given a $k$-space feature map $X = \{x^{\{1,1\}}, x^{\{1,2\}}, \ldots, x^{\{i,j\}}\}$, where $x^{\{i,j\}} \in \mathbb{R}^{C \times 1 \times 1}$ corresponding to channel $C$ and spatial location $(i, j)$ with $i \in [1, \ldots, W]$ and $j \in [1, \ldots, H]$, where $W$ and $H$ are width and height of a channel of k-space feature maps. The frequency-attention map $S_f \in \mathbb{R}^{W \times H}$ is learned with a convolutional operation $\circledast$:

$$S_f = \sigma(W_{\text{conv}} \circledast X), \tag{1}$$

where $W_{conv} \in \mathbb{R}^{C \times 1 \times 1}$ is convolutional weights to be learned and $\sigma$ is the sigmoid function. Each $S_f^{\{i,j\}}$ of the frequency-attention map represents the linearly combined representation of all channels for a spatial location $(i, j)$. Then the input feature map $X = \{X^1, \ldots, X^i\}$ is modulated by this frequency attention map $S_f$ to generate frequency-attention weighted features $Y_f = S_f X$.

### 2.2.2 Channel-attention layers

Since spatially adjacent slices may provide complementary information and data from different coils are sensitive to different regions of the object, we integrate multiple slices and data from multiple coils as a multi-channel input to the network. Instead of treating them equally, we adaptively learn weights for each channel using a channel-attention layer. Given k-space feature maps of multiple slices $X = \{X^1, ..., X^n\}$, where $n$ is the number of channels in $X$, we employ a simple gating mechanism with a sigmoid activation to learn a channel-attention map:

$$S_c = \sigma(W_{fc} \cdot [\|X^1\|_1, \|X^2\|_1, ..., \|X^n\|_1]), \tag{2}$$

where $W_{fc} \in \mathbb{R}^{n \times n}$ is parameters of a fully connect layer to be learned, $\sigma$ is the sigmoid function, and $\|.\|_1$ is used to squeeze each $X^i, i = 1, ..., n$ to yield a scalar value. Then the feature maps $X = \{X^1, ..., X^n\}$ are modulated by multiplying this channel attention map $S_c$ along the channel dimension to generate weighted features $Y_c = S_c \circ X$.

We use an element-wise Max-out operator to aggregate the weighted features obtained by the frequency-attention and the channel-attention layers:

$$Y = \max(Y_c, Y_f), \tag{3}$$

where $Y_c$ is the weighted features obtained by the channel-attention layer and $Y_f$ is the weighted features obtained by the frequency-attention features $Y_f$. Both frequency-attention and the channel-attention maps are learned with the entire network in an end-to-end learning fashion by minimizing the loss function.

### 2.2.3 Loss function

The network is optimized by minimizing a loss function $L$ defined as the mean square error (MSE) between the reconstructed image and its corresponding image obtained from fully sampled k-space data:

$$L = \|I^R - I^{FS}\|, \tag{4}$$

where $I^R$ is the reconstructed image and $I^{FS}$ is the image obtained from fully sampled k-space data.

## 2.3 Evaluation and ablation studies

### 2.3.1 Datasets

We evaluated our method using two publicly available datasets, including Stanford Fully Sampled 3D FSE Knee k-space Dataset (available at http://mridata.org/), and fastMRI Brain Dataset [26] (available at https://fastmri.org).

The Stanford dataset contains 20 cases of knee images, collected with 8 coils, repetition time (TR) = 1550 ms, and echo time (TE) = 25 ms. Each image is a 3D volume with 256 slices, slice thickness = 0.6 mm, field of view (FOV) = 160×128 mm², acquisition matrix size = 320 × 256, and pixel size = 0.5 × 0.5 mm². We randomly selected 15 cases as training data, one case as validation data, and the remaining four cases as test data.

The fastMRI brain dataset contains 4,478 cases, collected with 8, 12, 16, 20 or 24 coils. For the convenience of network training, we selected 570 cases collected with 16 coils from the dataset. Each case is a 3D volume with 16 slices. All images from this dataset were zero-padded to have the same size of 640 × 320. We randomly selected 500 cases as training data, 10 as validation data, and the remaining 60 as test data.

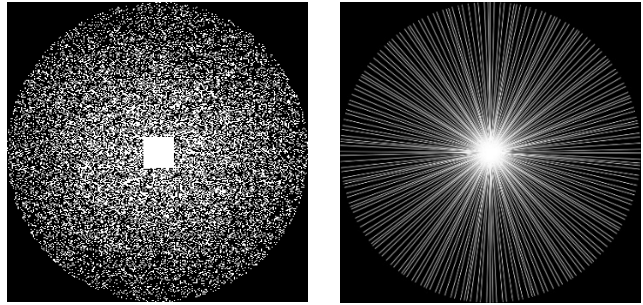

**Fig. 3.** Sampling masks. Left: Cartesian trajectory. Right: Radial trajectory.

The input *k*-space was undersampled using two different sampling strategies following [19], as illustrated in Fig. 3. A random Gaussian sampling with a 4x acceleration factor yields a Cartesian trajectory with a net acceleration factor R = 4. A radial sampling with 60 spokes (views) from full 256 spokes yields a Radial trajectory with a net acceleration factor R = 4. For the radial sampling, the non-uniform fast Fourier transform (NUFFT) [27, 28] was adopted to generate radial coordinate k-space data. And then Kaiser-Bessel gridding [29] was used to perform the regridding to the Cartesian coordinate with a square shape of 256 × 256 for both datasets.

### 2.3.2 Quantitative evaluation metrics

We adopted structural similarity (SSIM) index peak, signal-to-noise ratio (PSNR) and normalized mean square error (NMSE) as evaluation metrics. Particularly, $SSIM(\hat{v}, v) = (2\mu_{\hat{v}}\mu_v + c_1)(2\sigma_{\hat{v}v} + c_2)/[(\mu_{\hat{v}}^2 + \mu_{\hat{v}}^2 + c_1)(\sigma_{\hat{v}}^2 + \sigma_{\hat{v}}^2 + c_2)]$, $NMSE(\hat{v}, v) = \|\hat{v} - v\|_2^2/\|v\|_2^2$, and $PSNR(\hat{v}, v) = 10\log_{10}\max(v)^2/MSE(\hat{v}, v)$, where $\hat{v}$ is the reconstructed image, $v$ is the image reconstructed from fully-sampled *k*-space data, $\|.\|_2^2$ is a squared Euclidean norm, max $(v)$ is the largest value of $v$, $n$ is the number of entries of $v$, $\mu_{\hat{v}}$, and $\mu_v$ are the average value of pixel intensities in $\hat{v}$ and $v$ respectively, $\sigma_{\hat{v}}$ and $\sigma_v$ are their variances respectively, $\sigma_{\hat{v}v}$ is the their covariance, and $c_1 = k_1 L$ and $c_2 = k_2 L$ are two variables to stabilize the division with $L = \max(v)$, $k_1 = 0.01$, and $k_2 = 0.03$ [19]. The evaluation metrics were computed and averaged based on the center slices of reconstructed images to exclude slices that lie outside the anatomy. Particularly, we chose the middle 200 slices of the test cases to evaluate the image reconstruction performance for the Stanford dataset and all slices of the test cases for the fastMRI dataset.

The values of SSIM, PSNR, NMSE obtained by our method and those under comparison were quantitatively compared using Wilcoxon rank sum test.

### 2.3.3 Comparison with current state-of-the-art methods

We compared our method with a recently published *k*-space deep learning method [19], as well as an image domain deep learning method [30]. Both of these methods were built upon the same U-net architecture as illustrated in Fig. 1 (bottom). For fair comparison, we initially set the number of input slices to one. We also built a network with the number of input slices set to three using our method. The numbers of parameters of the different methods for the Stanford dataset and the fastMRI brain dataset are summarized in Table 1.

Table 1. Numbers of parameters of different methods.

| Methods | Number of input slices | Number of parameters | Stanford Dataset $N_{coil}$ = 8 | fastMRI Dataset $N_{coil}$ = 16 |
|---|---|---|---|---|
| Image-domain | 1 | 34.62M + 1216 $N_{coil}$ | 34.63M | 34.64M |
| *k*-space | 1 | 34.62M + 1216 $N_{coil}$ | 34.63M | 34.64M |
| ACNN-k-Space | 1 | 36.37M + 1216 $N_{coil}$ | 36.38M | 36.40M |
| | 3 | 36.37M + 3520 $N_{coil}$ | 36.38M | 36.43M |
| | $N_{slice}$ | 36.37M + 1152$N_{slice}N_{coil}$ + 64 $N_{coil}$ | | |

### 2.3.4 Visualization of attention maps

In order to understand how the frequency-attention layers modulate CNN features, we directly visualized the frequency-attention maps at different layers of the network.

In order to understand how the different channels of the multiple-channel input contribute to image reconstruction, we computed a response value for each channel as $\|grad(C)\|_1$, where $grad(C)$ is the gradient of a channel of the multiple-channel input once the network's weights were obtained. Since the channel-attention maps at other layers of the network are learned from combinations of data of multiple coils from spatially adjacent slices, we could not map the response values of different channels to the input slices or coils.

### 2.3.5 Ablation studies

We carried out ablation studies to investigate the effectiveness of frequency-attention layers and channel-attention layers. In all these experiments, we set the number of input *k*-space slice to three and used Cartesian sampling to undersample *k*-space data. We then evaluated how the image reconstruction performance changes with the number of input slices.

Due to the high computational cost, we did not evaluate all combinations of possible settings of our method. The ablation studies were carried out based on the Stanford dataset with Cartesian undersampling. Results of ablation studies on the fastMRI dataset are presented in a supplemental file.

The frequency-attention layers and channel-attention layers could learn attention weights in parallel or sequentially to module CNN features as illustrated in Fig. 4. To integrate features modulated by parallel attention layers, as illustrated in Fig. 4(a), we adopted an element-wise Max-out operator. To modulate features sequentially, the channel-attention and frequency attention layers could be integrated differently as illustrated in in Fig. 4b and Fig. 4c, respectively.

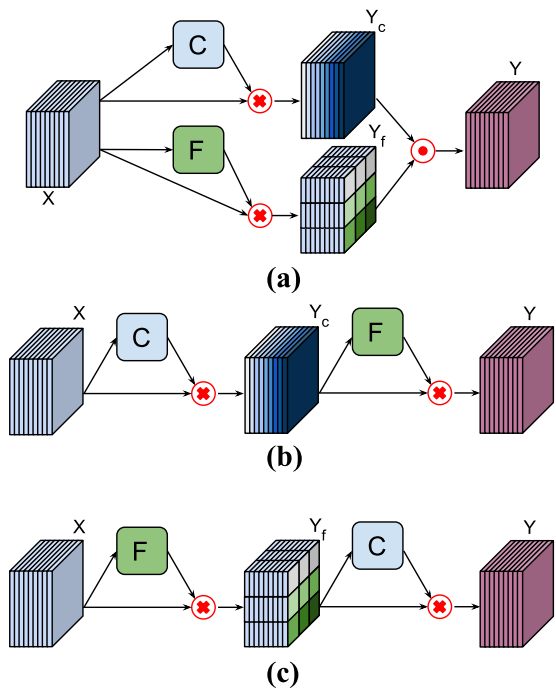

**Fig. 4.** Applications of attention layers. (a) The channel-attention and frequency-attention layers are applied in parallel. (b) The channel-attention layer is applied before the frequency-attention layer. (c) The frequency-attention layer is applied before the channel-attention layer.

*2.3.6 Implementation and computing cost*

We implemented all deep learning methods using PyTorch. Adam optimizer was used to train the network. Batch size was set to 16, number of training epochs set to 350, initial learning rate = $10^{-4}$ which gradually dropped to $10^{-5}$, and weight regularization parameter $\lambda = 10^{-4}$. All experiments were carried out on a Linux workstation equipped with 4 Titan XP GPUs with 12G memory. On a NVIDIA TITAN XP GPU, it took 50 milliseconds and 301 milliseconds for the deep learning models built by our method to achieve image reconstruction from undersampled Stanford k-space data (320×256×8) and fastMRI k-space data (640×320×16), respectively. It takes ~50 hours to train a deep learning model based on Stanford dataset and ~70 hours based on fastMRI dataset.

## 3   Experimental results

### 3.1 Image reconstruction performance of the methods under comparison

Image reconstruction performance comparisons are summarized in Tables 2 and 3 for Stanford and fastMRI datasets, respectively. These results demonstrate that our method consistently performs better than the existing methods (albeit by very small margins in some instances), in particular with an input of multiple slices (three in the current implementation if not specified otherwise). The *k*-space deep learning methods also had better performance than their counterpart in the image domain, consistent with previous findings [19].

**Table 2.** Performance of methods under comparison based on the Stanford dataset.

| Methods | Number of input slices | NMSE (× 10⁻³) | | | | PSNR | | | | SSIM (× 10⁻²) | | | |
|---|---|---|---|---|---|---|---|---|---|---|---|---|---|
| | | Cartesian | | Radial | | Cartesian | | Radial | | Cartesian | | Radial | |
| | | mean±std | p-value | mean±std | p-value | mean±std | p-value | mean±std | p-value | mean±std | p-value | mean±std | p-value |
| Image-domain | 1 | 14.77±0.01 | 9.55e-23 | 11.45±0.01 | 4.90e-15 | 34.25±1.82 | 9.55e-19 | 35.45±1.80 | 2.40e-44 | 89.95±0.03 | 8.47e-25 | 91.97±0.03 | 4.29e-45 |
| *k*-space | 1 | 11.98±0.01 | 2.07e-59 | 11.26±0.01 | 6.46e-33 | 35.27±1.89 | 7.43e-21 | 35.56±1.82 | 5.26e-67 | 92.05±0.02 | 4.08e-17 | 92.65±0.03 | 4.67e-30 |
| ACNN-k-Space | 1 | 11.55±0.01 | 7.35e-63 | 11.22±0.01 | 6.09e-23 | 35.45±1.81 | 2.09e-53 | 35.57±1.78 | 2.52e-63 | 92.15±0.03 | 6.33e-30 | 92.65±0.03 | 1.32e-27 |
| | 3 | **11.11**±0.01 | - | **10.90**±0.01 | - | **35.55**±1.86 | - | **35.63**±1.82 | - | **92.25**±0.03 | - | **92.73**±0.03 | - |

**Table 3.** Performance of methods under comparison based on the fastMRI brain dataset.

| Methods | Number of input slices | NMSE (× 10⁻³) | | | | PSNR | | | | SSIM (× 10⁻²) | | | |
|---|---|---|---|---|---|---|---|---|---|---|---|---|---|
| | | Cartesian | | Radial | | Cartesian | | Radial | | Cartesian | | Radial | |
| | | mean±std | p-value | mean±std | p-value | mean±std | p-value | mean±std | p-value | mean±std | p-value | mean±std | p-value |
| Image-domain | 1 | 20.43±0.02 | 1.81e-3 | 16.16±0.02 | 2.25e-3 | 38.25±3.23 | 6.05e-4 | 39.74±3.19 | 7.95e-7 | 94.11±0.05 | 1.27e-5 | 94.46±0.05 | 6.05e-4 |
| *k*-space | 1 | 19.67±0.02 | 3.34e-2 | 15.71±0.02 | 3.62e-6 | 38.60±2.97 | 3.39e-4 | 39.85±3.26 | 7.49e-3 | 94.58±0.05 | 5.27e-6 | 94.61±0.05 | 4.81e-9 |
| ACNN-k-Space | 1 | 18.80±0.02 | 2.32e-3 | 15.29±0.02 | 2.61e-2 | 38.95±3.20 | 2.57e-3 | 40.11±3.27 | 5.05e-e2 | 94.62±0.05 | 8.96e-5 | 94.86±0.05 | 5.57e-3 |
| | 3 | **18.30**±0.02 | - | **14.93**±0.02 | - | **39.16**±3.03 | - | **40.39**±3.30 | - | **95.02**±0.05 | - | **95.21**±0.05 | - |

Figure 5 shows representative results from the comparisons. Consistent with the quantitative results summarized in Tables 2 and 3, our method yields visually better results than the other two methods. In particular, the deep learning model in image domain yielded a slightly oversmoothed image around the region indicated by the arrow.

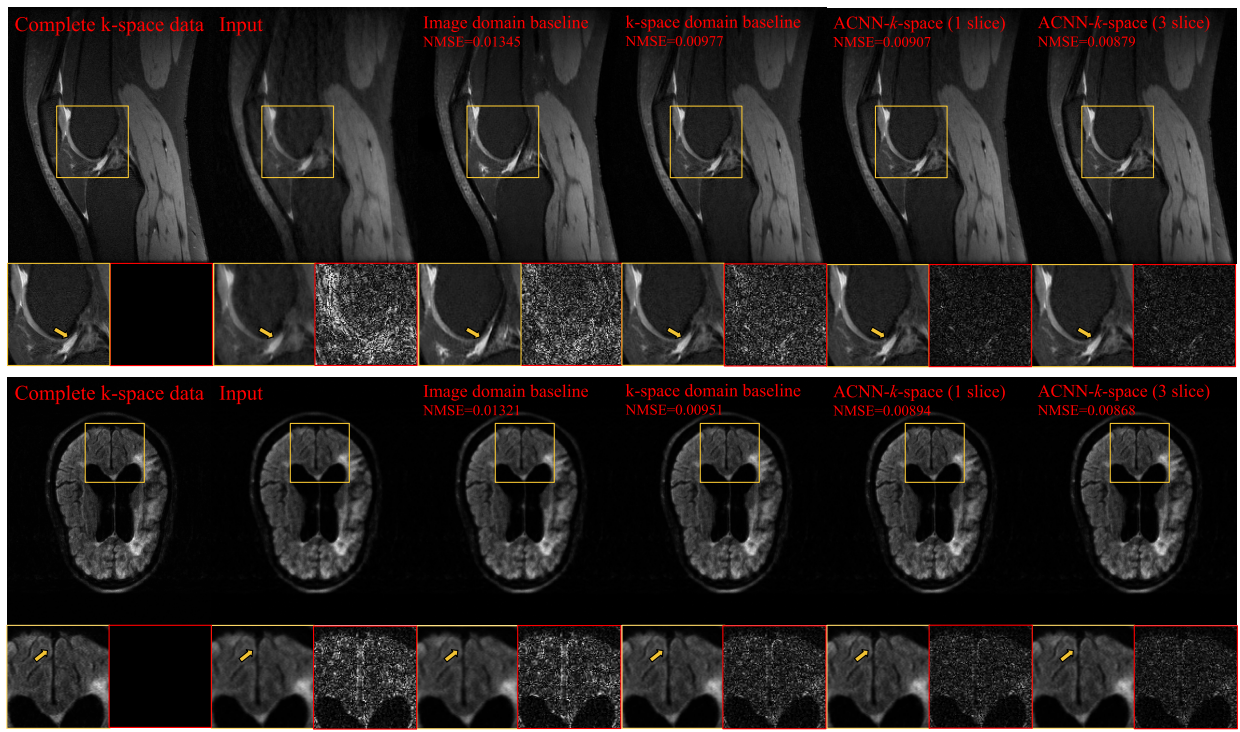

**Fig. 5.** Representative cases of the CNN reconstructions for the Stanford knee dataset (top row) and the fastMRI brain dataset (bottom row). Images reconstructed from fully sampled data (ground-truth) are also shown. Yellow and red boxes indicate the zoomed-in regions and difference images, respectively. Difference images were amplified 5-fold for Stanford cases and 10-fold for the fastMRI dataset.

## 3.2 Visualization of attention maps

Figure 6 shows representative frequency-attention maps learned by our method with the number of input slices set to three for image reconstruction from undersampled *k*-space data using Cartesian sampling based on the Stanford dataset (top row) and radial sampling based on the fastMRI dataset (bottom row). Not surprisingly, learned frequency-attention maps had varied weights at different spatial frequencies. It is worth noting that the last frequency-attention map largely complements the Cartesian sampling pattern to guarantee the network to generate a residual *k*-space interpolation map, which was added to the under-sampled *k*-space data through the residual connection (Fig. 1) to reconstruct the image.

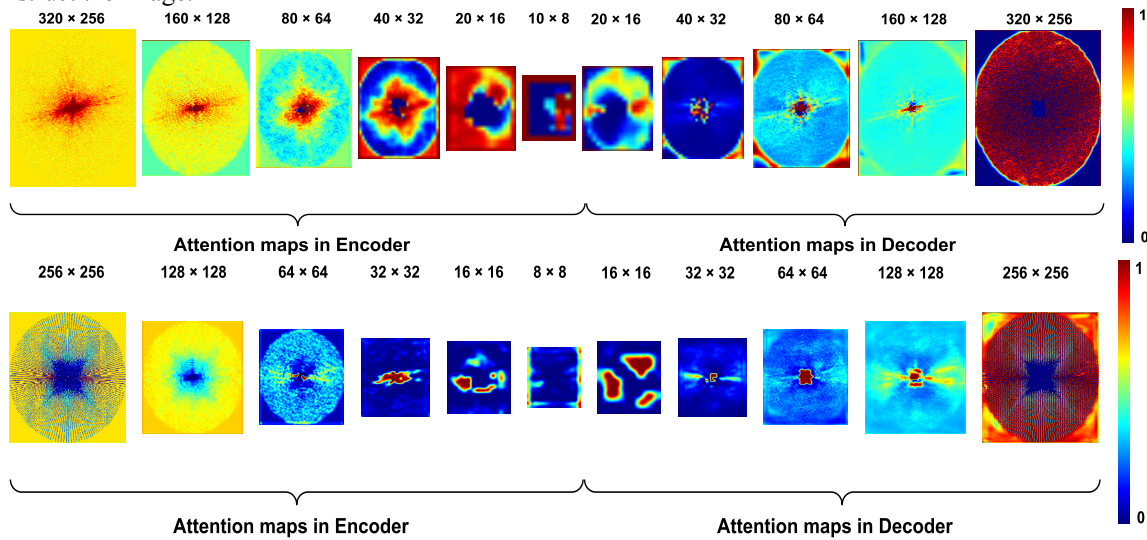

**Fig. 6.** Frequency-attention maps learned in the network for *k*-space data undersampled using Cartesian sampling (top) and radial sampling (bottom).

Figure 7 shows the response values of different channels grouped for individual slices for a representative test data of the Stanford knee dataset with the number of input slices set to 3, 5, 7, 9, 11, and 13, and using undersampled Cartesian *k*-space data. As expected, the channels (coils) of the center slice had the largest response values, and the response values decreased with distance from the center slice, indicating that the data of the center slice contribute most to image reconstruction while the influence decreases with distance from center.

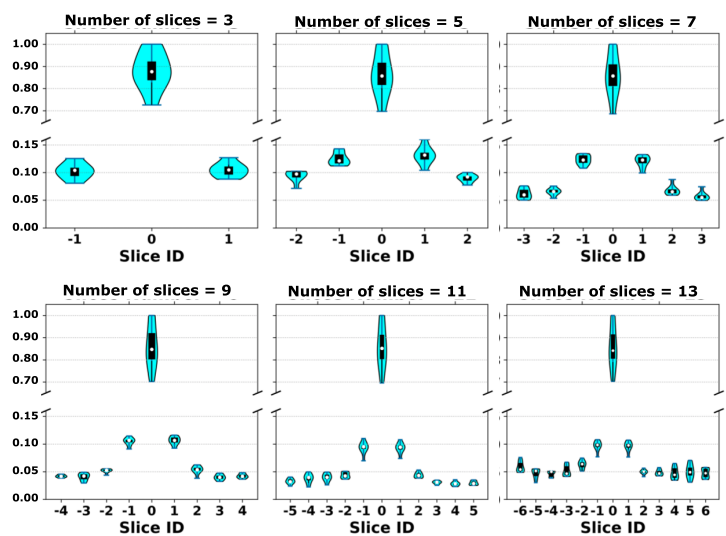

**Fig. 7.** Visualization of response values of different channels of a test case of the Stanford knee dataset. Response values were normalized by the maximum value of each slice group.

Figure 8 shows training loss functions for data undersampled with the Cartesian sampling and the Radial sampling on Stanford and fastMRI datasets, indicating that ACNN-k-Space works better on the training data with the Radial sampling than the Cartesian sampling although they have the same acceleration factor.

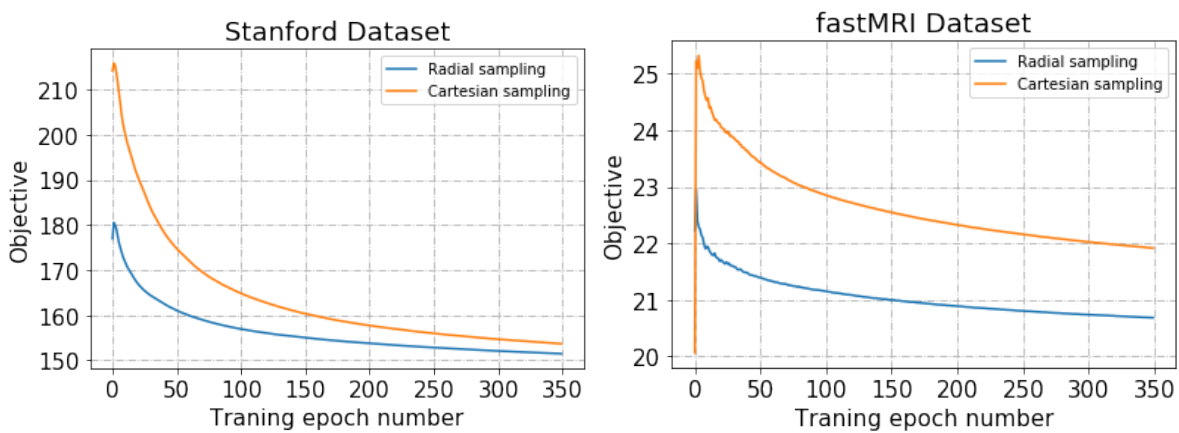

**Fig. 8.** Visualization of training loss functions for data undersampled with the Cartesian sampling and the Radial sampling. The objective value was obtained by normalizing the loss function value with the image size (width × length).

### 3.3 Ablation studies

As summarized in Table 4, our ACNN-k-Space method with both frequency- and channel-attention layers yielded the best performance measures, while methods utilizing one attention layer still outperformed those without the self-attention layers.

**Table 4.** Performance of the networks with different components based on the Stanford dataset.

| Methods | NMSE (× $10^{-3}$) | | PSNR | | SSIM (× $10^{-2}$) | |
|---|---|---|---|---|---|---|
| | Mean±std | p-value | Mean±std | p-value | Mean±std | p-value |
| Without self-attention | 11.95±0.01 | 7.35e-63 | 35.19±1.81 | 1.12e-59 | 92.04±0.02 | 2.02e-40 |
| Channel-attention alone | 11.52±0.01 | 9.74e-50 | 35.39±1.83 | 8.36e-48 | 92.15±0.03 | 8.07e-38 |
| Frequency-attention alone | 11.43±0.01 | 7.83e-37 | 35.43±1.87 | 5.32e-3 | 92.17±0.03 | 1.06e-17 |
| ACNN-k-Space | **11.11**±0.01 | - | **35.55**±1.86 | - | **92.25**±0.03 | - |

As summarized in Table 5, *k*-space deep learning method with F-C (frequency-attention layer followed by channel-attention layer) attention block performed better than C-F (channel-attention layer followed by frequency-attention layer) attention block, while the parallel frequency-attention and channel-attention layers yielded the best performance.

**Table 5.** Performance of different attention blocks on the Stanford dataset under Cartesian sampling mask.

| Methods | NMSE (× $10^{-3}$) | | PSNR | | SSIM (× $10^{-2}$) | |
|---|---|---|---|---|---|---|
| | Mean±std | p-value | Mean±std | p-value | Mean±std | p-value |
| C-F (Fig.4b) | 11.33±0.01 | 2.73e-67 | 35.47±1.81 | 2.73e-67 | 92.21±0.03 | 9.34e-47 |
| F-C (Fig.4c) | 11.26±0.01 | 1.07e-66 | 35.51±1.85 | 1.63e-66 | 92.24±0.03 | 2.34e-3 |
| Paralleled architecture | **11.11**±0.01 | - | **35.55**±1.86 | - | **92.25**±0.03 | - |

As summarized in Table 6, image reconstruction performance increased with the number of input slices. We expect that the method will reach its peak performance with a moderately large number of input slices. However, the computational cost will also increase with the number of input slices, while the incremental gain becomes increasingly smaller. Further, some datasets will be limited by the number of available slices.

**Table 6.** Performance of our method with different numbers of input slices based on the Stanford dataset.

| Number of input slices | NMSE ($\times 10^{-3}$) | | PSNR | | SSIM ($\times 10^{-2}$) | | Train times | Test time (each case) |
|---|---|---|---|---|---|---|---|---|
| | Mean±std | p-value | Mean±std | p-value | Mean±std | p-value | | |
| 1 | 11.55±0.01 | - | 35.45±1.81 | - | 92.15±0.03 | - | ~44 hours | ~49 ms |
| 3 | 11.11±0.01 | 7.35e-63 | 35.55±1.86 | 2.09e-53 | 92.25±0.03 | 6.33e-30 | ~50 hours | ~50 ms |
| 5 | 10.94±0.00 | 1.29e-64 | 35.57±1.84 | 5.11e-61 | 92.30±0.03 | 2.18e-42 | ~57 hours | ~58 ms |
| 7 | 10.76±0.00 | 2.00e-66 | 35.61±1.84 | 1.13e-61 | 92.35±0.03 | 8.75e-45 | ~65 hours | ~62 ms |
| 9 | 10.65±0.00 | 1.06e-65 | 35.64±1.84 | 1.15e-61 | 92.40±0.03 | 2.39e-45 | ~74 hours | ~65 ms |
| 11 | 10.64±0.00 | 2.88e-66 | 35.64±1.84 | 1.14e-61 | 92.45±0.03 | 9.74e-46 | ~84 hours | ~70 ms |

## 4 Discussion and Conclusions

We have developed a novel deep learning method for image reconstruction from undersampled *k*-space data. Our method is built upon a residual Encoder-Decoder network of CNNs to learn interpolation in *k*-space. Also, rather than learning the interpolation independently for each slice, we integrate complementary information of spatially adjacent slices as multi-channel input to the residual network to improve image reconstruction. We also adopt self-attention layers to effectively integrate complementary information of multiple slices and recognize distinctive contributions of *k*-space data at different spatial frequencies. Ablation studies and comparison experiments have demonstrated that our method could effectively reconstruct images from undersampled *k*-space data and achieved significantly better image reconstruction performance than state-of-the-art alternative techniques.

Our method is built upon a residual Encoder-Decoder network architecture with self-attention layers consisting of frequency-attention layers and channel-attention layers. As illustrated in Fig. 6, the frequency-attention maps had distinctive values at different spatial frequencies of the k-space data, indicating that the frequency-attention layers could modulate features learned by weight-sharing CNNs so that the deep learning models could model k-space data at low and high frequencies differently. As illustrated in Fig. 7, the channel-attention layers facilitate effective integration information from spatially adjacent image slices of the image slice under consideration, consistent with existing findings [9, 20]. The quantitative evaluation results summarized in Tables 2 and 3 have further demonstrated that the attention layers could effectively improve the image reconstruction performance compared with deep learning models without the attention layers with only a 5% increase in the number of model parameters.

Our network is built upon the standard residual Encoder-Decoder network architecture, which could be improved by adopting other network architectures, network blocks, and advanced learning strategies, such as Dense block [31], or instance normalization [32], in conjunction with advanced loss function [33]. It is noteworthy that a variety of deep-learning MR reconstruction methods have been developed recently, such as cascaded networks [9, 10], variational network [6], KIKI-network [34], RARE [35], and DeepcomplexMRI network [17]. Our method could serve a basic deep learning component to be integrated with these deep learning methods in a straightforward manner.

In conclusion, we have developed adaptive CNNs for k-space data interpolation which has achieved favorable performance compared with state-of-the-art deep learning methods.

# Supplementary data

Experimental results on the fastMRI brain dataset (available at https://fastmri.org).

As summarized in Table S1, our ACNN-k-Space method with both frequency- and channel-attention layers yielded the best performance measures, while methods utilizing one attention layer still outperformed those without the self-attention layers.

**Table S1.** Performance of networks with different components based fastMRI dataset under Radial sampling mask.

| Methods | NMSE ($\times 10^{-3}$) | | PSNR | | SSIM ($\times 10^{-2}$) | |
|---|---|---|---|---|---|---|
| | mean±std | p-value | mean±std | p-value | mean±std | p-value |
| Without self-attention | 15.47±0.02 | 1.90e-6 | 40.00±3.15 | 3.14e-9 | 94.83±0.05 | 2.67e-11 |
| Channel-attention alone | 15.25±0.03 | 5.61e-5 | 40.13±3.05 | 9.14e-7 | 94.94±0.05 | 4.19e-10 |
| Frequency-attention alone | 15.11±0.02 | 5.44e-6 | 40.23±3.02 | 3.37e-6 | 95.03±0.06 | 3.29e-12 |
| ACNN-k-Space | **14.93**±0.02 | - | **40.39**±3.25 | - | **95.21**±0.05 | - |

As summarized in Table S2, the *k*-space deep learning method with C-F (channel-attention layer followed by frequency-attention layer) attention block had the worst performance. F-C attention block had a better performance than C-F attention block. Our method with paralleled frequency-attention and the channel-attention layers had the best performance.

**Table S2.** Performance of different attention blocks on the fastMRI dataset under Radial sampling mask.

| Methods | NMSE ($\times 10^{-3}$) | | PSNR | | SSIM ($\times 10^{-2}$) | |
|---|---|---|---|---|---|---|
| | Mean±std | p-value | Mean±std | p-value | Mean±std | p-value |
| C-F (Fig.4b) | 15.02±0.02 | 5.23e-3 | 40.29±3.10 | 5.22e-6 | 95.13±0.05 | 7.31e-7 |
| F-C (Fig.4c) | 14.98±0.02 | 9.74e-3 | 40.35±3.25 | 8.51e-3 | 95.18±0.05 | 5.24e-8 |
| Paralleled architecture | **14.93**±0.02 | - | **40.39**±3.30 | - | **95.21**±0.05 | - |

As summarized in Table S3, image reconstruction performance increased with the number of input slices. We expect that the method will reach its peak performance with a moderately large number of input slices. However, the computational cost will also increase with the number of input slices, while the incremental gain becomes increasingly smaller. Further, some datasets will be limited by the number of available slices.

**Table S3.** Performance of our method with different numbers of the input slices based on fastMRI dataset.

| Number of input slices | NMSE ($\times 10^{-3}$) | | PSNR | | SSIM ($\times 10^{-2}$) | | Train time | Test time (each case) |
|---|---|---|---|---|---|---|---|---|
| | Mean±std | p-value | Mean±std | p-value | Mean±std | p-value | | |
| 1 | 13.54±0.02 | - | 40.68±3.44 | - | 95.39±0.05 | - | ~62 hours | ~58 ms |
| 3 | 13.28±0.02 | 6.13e-13 | 40.94±3.02 | 2.02e-9 | 95.77±0.05 | 4.75e-12 | ~70 hours | ~60 ms |
| 5 | 13.10±0.02 | 7.93e-8 | 41.02±3.01 | 4.55e-8 | 95.96±0.05 | 9.25e-9 | ~79 hours | ~64 ms |
| 7 | 12.98±0.02 | 9.01e-12 | 41.08±3.02 | 9.75e-9 | 95.98±0.05 | 6.06e-13 | ~89 hours | ~69 ms |
| 9 | 12.93±0.02 | 1.66e-11 | 41.10±2.98 | 7.20e-10 | 95.99±0.05 | 2.95e-11 | ~101 hours | ~73 ms |
| 11 | 12.89±0.02 | 1.32e-13 | 41.14±3.05 | 7.91e-10 | 96.00±0.05 | 7.08e-14 | ~113hours | ~78 ms |